\documentclass[useAMS,usenatbib]{mn2e}
\usepackage{amsfonts}
\usepackage{amsmath}
\usepackage{amssymb}

\def\bea{\begin{eqnarray}}
\def\ena{\end{eqnarray}}

\newcommand{\mr}[1]{\mathrm{#1}}

\usepackage[dvips]{graphicx}
\usepackage{amsmath,amssymb}
\usepackage{myaasmacros}
\usepackage{color}
\usepackage{multirow}

\title[Gamma-ray halo around the M31 galaxy]{Gamma-ray halo around the M31 galaxy as seen by the Fermi LAT}

\author[]{}

 \author[M. S. Pshirkov, V.V. Vasiliev \& K. A. Postnov]{M. S.
Pshirkov$^{1,2,3}$\thanks{E-mail:
 pshirkov@sai.msu.ru }, V. V. Vasiliev$^{4}$\thanks{E-mail:
 vasilyev@mpia.de }, K.A.Postnov$^{1}$\footnotemark[1]\thanks{E-mail:
 pk@sai.msu.ru}, \\
$^{1}$Sternberg Astronomical Institute, 
Lomonosov Moscow State University, 
Universitetsky prospekt 13, 119992, Moscow, Russia\\
$^{2}$Institute for Nuclear Research of the Russian Academy of
Sciences, 117312, Moscow, Russia\\
 $^{3}$Pushchino Radio Astronomy Observatory, 142290 Pushchino,
Russia\\
$^{4}$IMPRS Max Planck Institute for Astronomy,D-69117, Heidelberg,	
Germany 
 }

\begin{document}

\date{}

\pubyear{2015}

\maketitle

\label{firstpage}

\begin{abstract}
Theories of galaxy formation predict the existence of extended gas halo around spiral galaxies. If there are 10-100 nG magnetic fields at several ten kpc distances from the galaxies, extended galactic cosmic ray (CR) haloes could also exist. Galactic CRs can interact with the tenuous hot halo gas to produce observable gamma-rays. We have performed search for  a gamma-ray halo around the  M31 galaxy -- the closest large spiral galaxy. Our analysis of almost 7 years of the Fermi LAT data revealed the presence of a spatially extended diffuse emission excess around M31. The data can be fitted  using the simplest morphology of a uniformly bright circle. The best fit gave a 4.7$\sigma$ significance for a $0.9^{\circ}$ (12 kpc) halo  with a photon flux of $\sim (3.2\pm1.0)\times 10^{-9} ~\mathrm{cm^{-2}s^{-1}}$   and a  luminosity of  $(4.0\pm1.5)\times 10^{38} ~\mathrm{erg~s^{-1}}$ in  the energy range 0.3--100 GeV. Our results also imply a low level of the flux  from the disc of the M31 galaxy  $(3.3 \pm 1.0) \times 10^{-10}~\mathrm{cm^{-2}s^{-1}}$. The corresponding gamma-ray luminosity, $5\times10^{37} ~\mathrm{erg~s^{-1}}$ is several times smaller than the luminosity of the Milky Way. This difference could be explained by a lower star formation rate in M31: there are less CRs and the level of  the ISM turbulence is lower, which in turn leads to a shorter time of CR containment.
\end{abstract}




\section{Introduction}
\label{sec:intro}

Theories of galaxy formation predict the existence of extended haloes around  
spiral galaxies  due to gas inflow from their neighbourghood
\citep{White1978,Fukugita2006}. When falling,  this gas could be heated up to virial temperatures $10^6-10^7$ K, producing huge 
reservoirs of hot gas (coronae). There are several observational manifestations of these coronae: soft 
diffuse X-ray emission extending up to several ten kpc from the central galaxy 
\citep{Li2008}, absorption in O VII line \citep{Wang2005,Bregman2007}, 
distortions in the shape of gas clouds \citep{Westmeier2005} and stripping of 
gas in the satellite galaxies by the ram-pressure of the halo gas 
\citep{Blitz2000}, see \citep{Putman2012} and references therein for an 
extensive review. The existence of such a hot halo around the Milky Way is 
established by several different methods \citep{Miller2013}. 

Extragalactic hot haloes are extremely elusive: they have been observed  only in 
several normal galaxies without star-formation bursts \citep{Bogdan2013}. However, recently the implementation of  stacking technique   showed 
that there is a statistically significant excess in the extended X-ray signal 
from some normal late-type galaxies as well \citep{Anderson2013}, suggesting that even if the 
haloes are individually undetectable at the present level of sensitivity, they  
still could be discovered by analysis of groups of objects.

The Milky Way and other disc galaxies can also be immersed into extended cosmic rays (CRs) haloes. This idea was thoroughly investigated in \citep{DePaolis1999,Feldmann2013}. 
It is well known that the Milky Way is not a perfect calorimeter for CRs: they rather quickly, on time scales of 10-20 Myr, escape from dense regions of the Galaxy, losing only minor part of their energy in interactions with the 
interstellar medium \citep{Strong2007, Strong2010}. However, if strong enough 
magnetic fields exist far away from the central regions of the Galaxy, these CRs 
would not go directly away to the intergalactic space, but would be instead retained in the magnetized galactic halo for a considerable time. Magnetic fields  10-100 times as weak as the galactic ones  
$\mathcal{O}(\mu$G) could be sufficient to contain these CRs for the cosmological time.
Wandering CRs would interact with tenuous ($\sim10^{-4}~ \mathrm{cm^{-3}}$) hot 
plasma producing gamma-rays via  pionic channel. Estimates  show that the gamma-ray luminosity of such a  halo could be around $10^{39}~\mathrm{erg~s^{-1}}$ 
at energies above 100 MeV \citep{Feldmann2013}. The size and shape of the halo cannot be firmly 
established and depend crucially on the propagation properties of CRs. The halo 
'half-light' radius is estimated to be 20-40 kpc \citep{Feldmann2013}.

The contribution of the CR halo around our Galaxy to the isotropic gamma-ray background can be as high as 10$\%$, and it is difficult to disentangle it from 
the truly extragalactic component. However, such haloes can be searched for around other spiral galaxies. The most natural target is the halo 
around the nearby M31 (Andromeda) galaxy. With the expected angular size of several degrees and a gamma-ray luminosity of  
$\sim10^{39}~\mathrm{erg~s^{-1}}$, such a halo could be detected by the Fermi LAT even from the Earth-M31 distance of $>700$ kpc. The presence of a hot gas around M31, which is essential for the gamma-ray emission from the CR halo, was recently demonstrated by the discovery of certain absorption features in UV-spectrum of quasars projected on the sky close to the galaxy \citep{Rao2013,Lehner2014} and distortions in the observed CMB spectrum in the vicinity of M31 due to interference from the halo gas \citep{DePaolis2014}.

The paper is organized as follows: in  Section II we describe the data and method of data analysis, Section III contains our results, and  summary and discussion
are in Section IV.

\section{Data and data analysis}
\label{sec:data}

In our analysis we have used 81 months of Fermi LAT data collected  since 2008 Aug 04 ( MET =239557417 s)  
until 2015 Jul 06 (MET=457860004 s).  We have selected events that belong to the "SOURCE" 
class in order to have a sufficient number of events without losing in their 
quality. The   PASS8\_V2 reconstruction and v10r0p5\footnote{http://fermi.gsfc.nasa.gov/ssc/data/analysis/software/} version of the Fermi 
science tools  was used.
As the expected signal is weak and diffuse, we have selected events 
with energies larger than 300 MeV, because at lower energies the Fermi LAT point 
spread function (PSF) quickly deteriorates. 
Usual event quality cut, namely that the zenith angle should be   less than 
$100^{\circ}$ (which is sufficient at these energies)   has been 
imposed.

Smaller PSF allowed us to use smaller region of interest (RoI) as well -- we took a circle of 10 degrees around the centre of the M31 galaxy 
($\alpha_{J2000}=10.6846^{\circ}, \delta_{J2000}=41.2692^{\circ}$).  The data were analysed using the binned maximum likelihood approach \citep{Mattox1996} implemented in the \textit{gtlike} utility, in which two  model hypotheses were compared by  their maximal   likelihoods with respect to  the observed photon distribution. The null hypothesis   does not include the halo,  the alternative hypothesis adds the halo to the list of sources of the null hypothesis. 

The source model includes 26  sources from the 3FGL catalogue \cite{3FGL}, the latest galactic interstellar
emission model gll$\_$iem$\_$v06$\_$rev1.fit, and the isotropic spectral template iso$\_$source$\_$v06.txt\footnote{http://fermi.gsfc.nasa.gov/ssc/data/\\access/lat/BackgroundModels.html}. Parameters of these sources were allowed to change. We also included additional 69 point-like gamma-ray emitters from the 3FGL catalogue between $10^{\circ}$ and $15^{\circ}$ from the RoI center with their parameters  held fixed.

The  M31 galaxy itself was modeled as an extended source based on the IR observations 
\citep{iras} ($100\mu$m normalized IRIS map) following the prescriptions of the Fermi LAT collaboration 
\citep{m31_fermi}.

Finally, extended halo spatial templates were inserted into the source model. We 
have used the simplest spatial models -- uniformly bright circles of 
different radii (from 0.1 to 5.0 degrees with 0.1 degree step). Of course, it is not a 
realistic model, because some decrease in surface  brightness towards the
outer halo regions can be expected. On the other hand, scarcity of the data used 
justifies this simple approach -- a more sophisticated  model would inevitably involve a 
 larger number of parameters, which would make fitting much harder and 
would dilute any obtained significance as well.

The M31 galaxy and the  halo were described by a simple power-law model:
\begin{equation}
dN/dE = N_{0} (E/E_{0})^{-\Gamma}
\end{equation}

The normalization $N_{0}$ and spectral index $\Gamma$ were allowed to vary during the likelihood optimisation, while the energy scale $E_{0}$ was fixed at 1 GeV. 

The evidence of the detection of gamma-ray signal from the halo was evaluated in terms of a likelihood ratio test statistic:
\begin{equation}
TS = - 2 \ln \frac {L_{max,0}} {L_{max,1}}
\end{equation}
where $L_{max,0}$ and $L_{max,1}$ are maximum likelihood values obtained from the observed data fit using null and alternative hypothesis, respectively. If the \textit{alternative} hypothesis is true, then $\sqrt{TS}$ is approximately equivalent to the source detection significance.

\section{Results}

Firstly, we have performed our analysis without additional source. The M31 galaxy was modelled in two different ways: as a point-like source or as an extended object (the IRAS template). 
The extended template for the M31 galaxy fits the data considerably better than the simple point-like source ($TS_{\mr{ext}}=79$, $TS_{\mr{ps}}=62.3$). The galaxy has a soft spectrum with photon index $\Gamma=2.40\pm0.12$ and the  flux $F=(2.6\pm0.4)\times10^{-9} \mathrm{ph ~cm^{-2}~ s^{-1}}$ in the 0.3-100 GeV energy range. The spectrum is even softer if the galaxy is modelled as a point-like source: $\Gamma=2.64\pm0.15$ with the  photon flux $F=(1.9\pm0.3)\times10^{-9} \mathrm{ph ~cm^{-2}~ s^{-1}}$.

\begin{figure}
\begin{center}
\includegraphics[scale=0.3, angle=270]{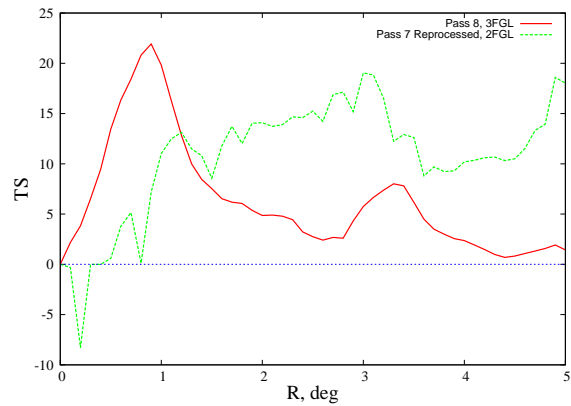}
\end{center}
\caption{ $TS (R_{\mr{halo}})$ curves: an earlier version (65 month of data, Pass7 Reprocessed events and 2FGL catalogue) is shown for  comparison. The $TS (R_{\mr{halo}})$ curve is much smoother when the latest version of event reconstruction and 3FGL source catalogue are used.}
\label{fig:graph}
\end{figure}

\begin{figure}
\begin{center}
\includegraphics[scale=0.3, angle=270]{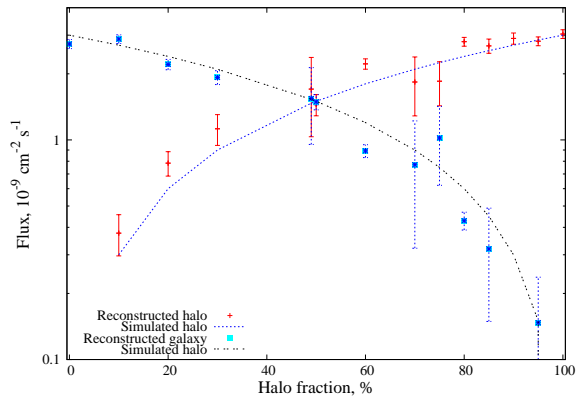}
\end{center}
\caption{ Comparison of the reconstructed and simulated fluxes from the disc and halo components for different halo fractions in the total flux. }
\label{fig:flux_sim}
\end{figure}

 The results of fitting with additional halo component are presented in  Fig.~\ref{fig:graph}. The fit 
 quality improvement can be easily seen.
The highest statistical significance $TS=22$ was obtained for a halo with radius  $R_{\mr{halo}}=0.9^{\circ}$, 
corresponding to a linear size of $\sim 12$ kpc.
The photon flux  from the extended halo and the 0.3-100 GeV luminosity obtained from the fit are $\sim (3.2\pm 1.0)\times 10^{-9}$ $\mathrm{cm^{-2}s^{-1}}$ and $(4.0\pm1.5)\times 10^{38}~\mathrm{erg~s^{-1}}$, respectively, adopting the distance $d=780$~kpc. The spectral index is found to be rather soft: $\Gamma=2.30\pm0.12$.  A marginal improvement ($
TS \sim 8$ )  could be also achieved by adding a 3-degree halo ($\sim35~$ kpc). Note that in the case of the small halo the total signal from the M31 region is dominated by the halo rather than the galaxy disc: 
the flux from the disc is found to be $\sim 10\%$ of the total flux $(\sim (3.3 \pm 1.0) \times 10^{-10}~\mathrm{cm^{-2}s^{-1}})$. This fact suggests that the IR-based template cannot 
fully trace the gamma-ray emission, and this emission is far more extended than the template size.  
To find how the observed gamma-ray flux from the region is shared between the two components,   
we have simulated events in the energy range  0.3--100 GeV for the relevant time span (71 months) and the RoI described above. The model included the galactic and isotropic backgrounds, 24  point-like sources from the 3FGL catalogue,  the M31 galaxy disc (taken in the form of the IRAS 100$\mu$m template). Spectral parameters and photon fluxes for the point-like sources were taken from the 3FGL catalogue. The recommended values were taken for the isotropic  and galactic backgrounds   fluxes\footnote{http://fermi.gsfc.nasa.gov/ssc/data/analysis/scitools/help/\\	gtobssim.txt}. We have fixed the total flux from the halo and disc components to a fiducial value $3.0\times	10^{-9}~\mathrm{cm^{-2}s^{-1}}$  and performed simulations, gradually  changing the halo contribution from 0 to 100 $\%$. The results are presented in Fig.\ref{fig:flux_sim}. First of all, note that there is no leakage of the disc photons to the halo -- when the fraction of the simulated halo photons is low, the results of the corresponding fit immediately show this. The same is true for the disc component as well. Clearly, this method can  effectively separate the two components. 
   
   The actual flux from the M31 galaxy disc is low, $<4\times10^{-10}~\mathrm{cm^{-2}s^{-1}}$, therefore its luminosity in the 0.3-100 GeV energy range is rather modest: $5\times10^{37}~\mathrm{erg~s^{-1}}$. This value is several times as small as that of the Milky Way \citep{Strong2010}\footnote{This luminosity also includes some contribution from the Milky Way halo, so the direct comparison is not straightforward}. This difference could be explained by low level of the  star formation rate in  M31, which is smaller than the corresponding Galactic rate  by a factor of  4-5 \citep{Kennicutt2012,Ford2013}. The lower star formation rate not only implies less energy in the form  of cosmic rays that eventually produce the observed gamma-rays, but also means that the level of turbulence in the ISM of M31 is lower, which in turn leads to a rapid escape of CRs from the M31 disc region.

\begin{figure}
\label{fig:tsmap}
\begin{center}
\begin{picture}(220,170)
{\includegraphics[width=.8\columnwidth]{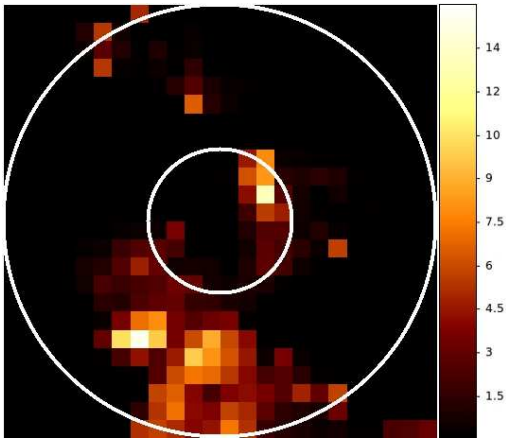}}
\end{picture}
\end{center}
\caption{ The $TS$ map with the IRAS template for the M31 disc. A complex extended structure around the galaxy is clearly seen. 1- and 3- degree radius white circles are shown for convenience.}
\end{figure}

\begin{figure}
\begin{center}
\begin{picture}(220,150)
{\includegraphics[width=.8\columnwidth]{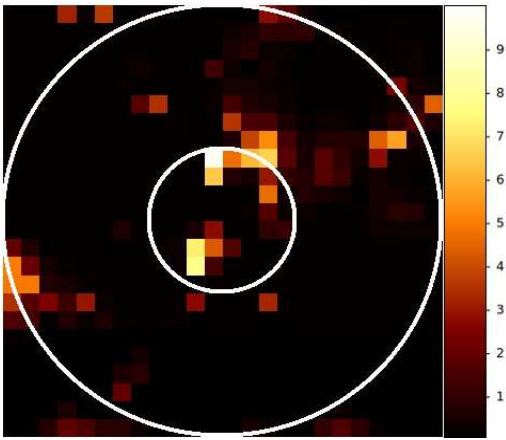}}
\end{picture}
\end{center}
\caption{The $TS$ map of simulated data including 0.9-degree halo with the photon flux $F_{\mr{0.3-100 GeV}}=1.5 \times 10^{-9}~\mathrm{cm^{-2}s^{-1}}$. 
A bright spot with $TS\sim10$ that emerged by chance is seen. There are  no clear signs of any extended structure beyond the $1^{\circ}$ radius.}
\label{fig:tsmap_sim}
\end{figure}

 To exclude possible systematic and instrumental effects, which could affect our results, we have performed several additional tests:

(i) In order to check whether the $TS$ increase corresponding to the 0.9-degree halo was caused by some unidentified point-like sources, not included into the 3FGLcatalogue, we have calculated the $TS$ map using the $gttsmap$ utility (see fig.~\ref{fig:tsmap}). A TS exccess at about 0.9 degree from the center of the galaxy  with the galactic coordinates ($l=120.58^{\circ}, b=-21.17^{\circ} $) emerges that could be ascribed  to FSRQ B3  0045+013. However, even after adding  the source with these coordinates into our source model, the $TS$ of the halo decreased only from  22 to 15 (the $TS$ for this source was 11.2), thus the whole increase could not be attributed to this  source alone. Alternatively,  this $TS$ excess could be produced by an inhomogeneity in the M31 halo. The plausibility of this scenario is also confirmed by the inspection of the TS maps of several simulated haloes -- they are far 
from being smooth and uniform, but rather consist of several random knots that could have $TS>10$ (see Fig. \ref{fig:tsmap_sim}).

(ii) We have also checked that the smallness of our RoI does not considerably affect our analysis: we have performed the data analysis using a larger circle with 15$^{\circ}$ radius. The $TS$ values from the halo remained essentially unchanged.

\begin{figure}
\label{fig:template_pic}
\begin{center}
\includegraphics[scale=0.3]{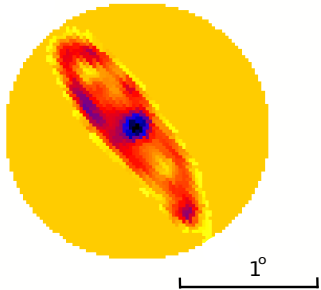}
\end{center}
\caption{Templates of the M31 galaxy disc and the halo with radius R=1.0$^{\circ}$.}
\end{figure}

\section{Summary and conclusions}
\label{sec:conclusions}

Using almost 7 years of the Fermi-LAT data, we have performed searches for an extended gamma-ray halo at energies above 300 MeV around the closest large spiral galaxy, M31. Such a gamma-ray halo could have appeared as a result of interactions of CRs  from the M31 galaxy with gas in its halo. 
We find that the Fermi-LAT data suggest the presence of a spatially extended  gamma-ray excess around  M31.  The data can be described  using the simplest morphology of a uniformly bright circle. The best fit gave $\sim 4.7\sigma$ significance for a $0.9^{\circ}$ radius (12 kpc) halo with  the photon flux $\sim (3.2\pm1.0)\times 10^{-9} ~\mathrm{cm^{-2}s^{-1}}$   and luminosity  $(4.0\pm1.5)\times 10^{38} ~\mathrm{erg~s^{-1}}$ in  the energy range 0.3--100 GeV.   The presence of such a halo compellingly shows that a substantial magnetic field ($>100$ nG) should extend around M31 up to at least  10 kpc. 
This excess could indicate the presence of a compact CR halo of 10-15 kpc in  radius, similar to the halo that is indirectly observed  around the Milky Way.
Independent  observational checks of such a circumgalactic magnetic fields could be done, for example, by analysis of the Faraday rotation measures of background extragalactic sources (see, e.g.,  \citep{Pshirkov2011}). In addition, synchrotron emission from secondary leptonic CR component could contribute at radio-frequencies $\le100$ MHz. Finally, gamma-ray signatures of dark matter particle annihilations (or decays) around M31 may be expected \citep{Baltz2008,Dugger2010}.

Our results also imply a low level of the gamma-ray flux  from the M31 galaxy disc --  $(3.3 \pm 1.0) \times 10^{-10}~\mathrm{cm^{-2}s^{-1}}$. The corresponding gamma-ray luminosity, $5\times10^{37} ~\mathrm{erg~s^{-1}}$, is several times smaller than the corresponding gamma-ray luminosity of the Milky Way. This difference could be explained by a lower star formation rate in M31: there are less CRs and the level of  the ISM turbulence is lower, which in turn leads to a shorter time of containment.

Past activity of the M31 galaxy could have been responsible for  the  complex structure of the $TS$ excess at several degrees scale -- see the example of the Fermi bubbles in the Galactic center. The SMBH in the center of M31 that is almost by two orders of magnitude more massive than SMBH in the Milky Way  and could have injected much more energy in the form of cosmic rays in the circumgalactic space. Future observations, including at  energies $>100$ GeV \citep{VERITAS, HAWC} would certainly clarify this issue.

\section*{Acknowledgements}
The work  was supported by the Grant of the President of Russian Federation MK-2138.2013.2, MK-4167.2015.2 and RFBR grant 14-02-00657.  M.P. acknowledges the fellowship of the Dynasty foundation. The authors want to thank Dmitry Prokhorov and Igor Moskalenko for fruitful discussions. The analysis is based on data and software provided by the Fermi Science Support Center (FSSC). The numerical part of the work was done at the computer cluster of the Theoretical Division of INR RAS and cluster of the SAI MSU. 
This research has made use of NASA's Astrophysics Data System, NASA/IPAC Extragalactic Database (NED) which is operated by the Jet Propulsion Laboratory, California Institute of Technology, under contract with the National Aeronautics and Space Administration, and the SIMBAD database, operated at CDS, Strasbourg, France.


\bibliographystyle{mn2e}

\end{document}